\documentclass[usegraphix,usenatbib]{mn2e}
 
\include{defn}

\usepackage{amsmath}
\usepackage{amssymb}
\usepackage{times}
\usepackage{graphicx}
\usepackage{subfigure}
\usepackage{epsfig}
\usepackage{txfonts}

\def\half{{\frac{1}{2}}}
\def\threeh{{\frac{3}{2}}}

\def\Rc{{R_{\rm c}}}
\def\psie{{\psi_{\rm env}}}
\def\Fe{{F_{\rm even}}}

\begin{document}

\pagenumbering{arabic}

\title[Extremely Flat Haloes and the Shape of the Galaxy ]
  {Extremely Flat Haloes and the Shape of the Galaxy}

\author[Evans \& Bowden]
  {N.W. Evans$^1$\thanks{E-mail:nwe. adb61@ast.cam.ac.uk},
   A. Bowden$^1$
 \medskip
 \\$^1$Institute of Astronomy, University of Cambridge, Madingley Road,
       Cambridge, CB3 0HA, UK}

\maketitle

\begin{abstract}
We present a set of highly flattened galaxy models with asymptotically
constant rotation curves. The mass density in the equatorial plane
falls like (distance)$^{-1}$ at large radii.  Although the inner
equidensity contours may be spherical, oblate or prolate, the outer
parts are always severely flattened. The elongated shape is supported
by rotation or tangential velocity anisotropy. The models are
thickened Mestel discs, and form a previously undiscovered part of the
Miyamoto \& Nagai sequence of flattened galaxies. The properties of
the models -- axis ratios, velocity dispersions, streaming velocities
and distribution functions -- are all discussed in some detail.

We pose the question: are extremely flattened or disk-like haloes
possible for the Milky Way galaxy? This has never been examined before,
as very flattened halo models were not available. We fit the rotation
curve and the vertical kinematics of disc stars in the solar
neighbourhood to constrain the overall shape of the Galaxy. Denoting
the ratio of polar axis to major axis by $q$, we show that models with
$q\lesssim 0.57$ cannot simultaneously reproduce the in-plane and
out-of-plane constraints. The kinematics of the Sagittarius galaxy
also strongly disfavour models with high flattening, as the orbital
plane precession is too great and the height reached above the
Galactic plane is too small. At least for our Galaxy, the dark halo
cannot be flatter than E4 (or axis ratio $q \sim 0.57$) at the Solar
circle.  Models in which the dark matter is accounted for by a massive
baryonic disc or by decaying neutrinos are therefore ruled out by
constraints from the rotation curve and the vertical kinematics.
\end{abstract}


\section{INTRODUCTION}

It is well-known that the Galaxy's rotation curve is flat to a good
approximation at large radii~\citep{So09}. As such, it is of course
typical of large spiral galaxies~\citep{So01}.

Equally well-known is the fact that -- under the assumption of
Newtonian gravity -- this implies the existence of ample dark matter
at large radii. If the dark matter is distributed in a roughly
spherical manner, then its density must fall like (distance)$^{-2}$ at
large radii to generate a flat rotation curve. This is of course
exemplified by models such as the isothermal sphere and its
cousins~\citep[see e.g.][chap. 4]{BT}. If the dark matter is
distributed in an extremely flat or disc-like configuration, then
its density must fall like (distance)$^{-1}$. This is the case for the
razor-thin Mestel (1963) disc.

The shapes of haloes -- whether they are roundish or flattish --
remain largely unconstrained. There are a variety of techniques by
which shapes can be inferred, such as analysis of the stellar
kinematics, the flaring of the neutral HI gas, the kinematics of
streams, the isophotes of the X-ray emitting gas, and studies of warps
and polar rings. However, all the techniques are indirect and involve
some assumptions (such as tilt of the velocity ellipsoid or
hydrostatic equilibrium) which may not be generally valid.  If $q$ is
the ratio of polar axis of the halo density contours to major axis,
then values in the range 0.1 (highly flattened and oblate) to 1.4
(prolate) have been reported for nearby galaxies. In particular, there
are claims of highly flattened haloes with axis ratios $q\sim 0.2$ for
NGC 891~\citep{Be97} and NGC 4244~\citep{Ol96} derived form analysis
of the flaring HI gas layer.

For the Milky Way Galaxy, the halo shape has been measured by at least
four different methods (stellar kinematics, alignment of velocity
ellipsoid, flaring and warping of neutral gas layer and kinematics of
the Sagittarius (Sgr) stream). The reported flattenings $q$ cover a
disarmingly wide range of values, from $q =0.4$~\citep[Jeans modelling
  of SDSS stars by][]{Lo12} to $q=1.7$~\citep[Sgr stream
  modelling by][]{He04}, with a sprinkling of values centered on
$q\approx 1$ or nearly spherical~\citep{Fe06,Sm09} for good measure.
Triaxiality is a further possible complication and there have also
been claims of halo triaxiality from modelling of the disruption of
the Sgr galaxy~\citep{La10,Deg13}. The only real conclusion is that systematic
effects in the various measurements techniques have been not yet been
properly understood.

Given the observation picture is so muddled, can we seek quite
guidance from theory? If the dark matter is a cold and weakly
interacting, then dissipationless simulations suggest that haloes
should be triaxial, almost prolate~\citep[e.g.,][]{Al06}. More
recently, dissipative simulations of galaxy formation in the
$\Lambda$CDM universe have become standard and they suggest that halo
shapes are typically mildly oblate. The ratio of short to long axis is
$q \approx 0.8-0.9$ though with some scatter~\citep[see
  e.g.,][]{Ab10, De11, Ze12}. Other dark matter candidates give rather
different predictions. Highly flattened ($q \sim 0.2$) haloes are
necessary if the dark matter is composed of decaying
neutrinos~\citep{Sc93,Sc99}. There also remains the possibility that
the flatness of rotation curves is baryonic in origin, either due to
cold molecular gas~\citep{Pf94a, Pf94b} or opaque atomic
gas~\citep{Kn01}.  In fact, heavy baryonic discs do have some
advantages, as they circumvent the disc-halo conspiracy and provide
simple explanations of the known disc scaling laws, as pointed out
recently by~\citet{Da12}.

Here, we ask the question: can highly flattened or disky dark haloes
be ruled out for the Milky Way galaxy? This question does not appear
to have been addressed before, probably because there are no flat
rotation curve models available in the literature that actually span
the gamut of shapes from prolate to very severely flattened. To solve
our problem, we have to first develop the potential theory of disky
dark haloes, then show such models have physically realistic,
everywhere positive distribution functions and finally examine
whether they can reproduce the kinematics of stars in the Milky Way.

We introduce our family of extremely flat galaxy models in Section
2. They fatten up the infinitesimally thin Mestel discs, and provide a
sequence of models which are always asymptotically very flat, but may
be prolate, oblate or spherical in the center. At large radii, the
three dimensional density falls like (distance)$^{-1}$, as befits
their disc-like nature.  Our models belong to a famous family devised
by \citet{Mi75} and \cite{Na76}, but eluded their original
investigations -- no doubt because their work actually predated the
discovery of flat rotation curves in the late 1970s.  In fact, as the
missing models have asymptotically constant rotation curves, they may
be the most useful of all, especially for modelling highly flattened
galaxies.

Section 3 discusses the properties of the models, and provides
solutions of the Jeans and collisionless Boltzmann equations. Given
that the models are very flattened, we construct the distribution
function to ensure that they are physical. This cannot be taken for
granted, as for example, the familiar flattened logarithmic model does
not have an ellipticity greater than E3~\citep[see e.g.,][]{Eva93,BT}.
Then, in Section 4, we apply these models to constrain the shape of
the Milky Way halo. Using the vertical kinematics at the solar
neighbourhood, together with the rotation curve, we argue that the
shape of the Milky Way halo at or close to the Sun has to be rounder
than $q =0.57$. This therefore rules out theories in which the flatness
of the rotation curve is generated by baryonic material confined to a
disc-like configuration.

\begin{figure}
    \centering
    \includegraphics[width=2.5in]{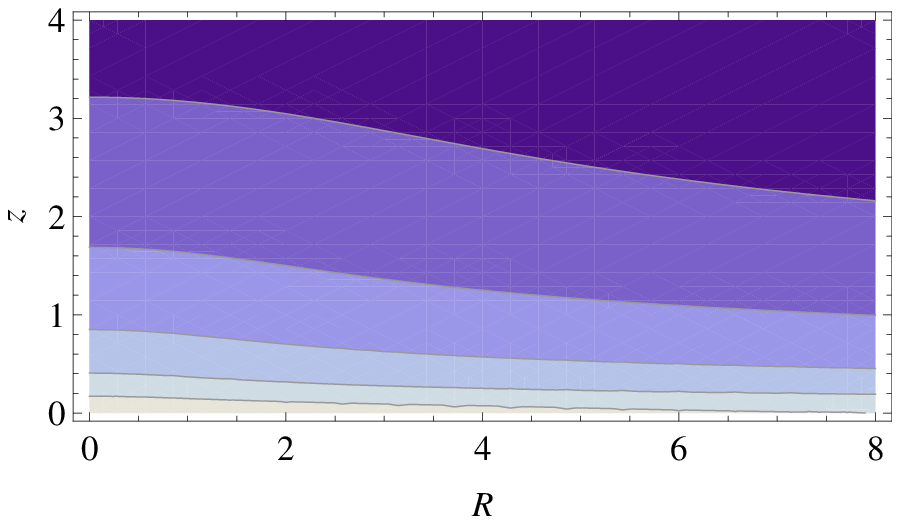}
    \includegraphics[width=2.5in]{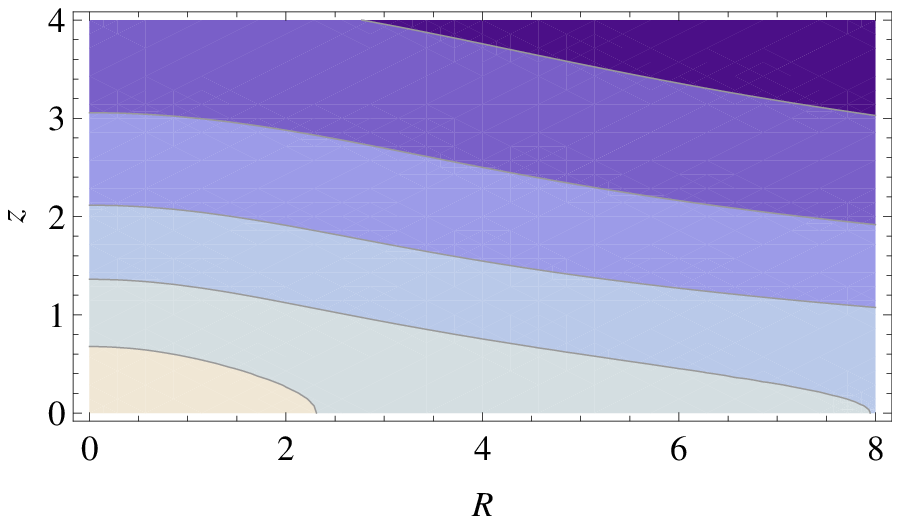}
    \includegraphics[width=2.5in]{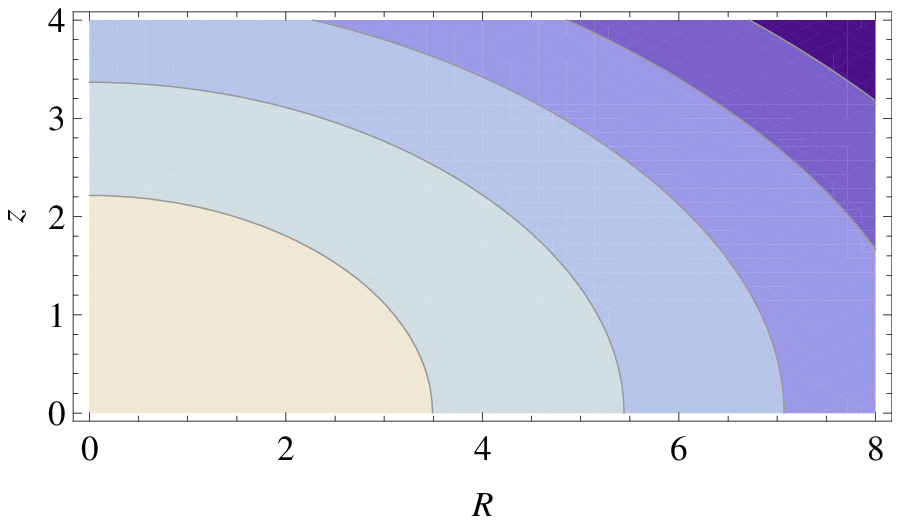}
    \caption{Logarithmic density contours in ($R,z$) plane for the
      missing model (\ref{eq:newdens}) with $b/a = 0.1, 1$ and 10. All
      the models are flattened, with $b/a \rightarrow 0$ being the
      razor-thin (Mestel) disc limit.}
    \label{fig:densfigure}
\end{figure}
\begin{figure}
    \centering
    \includegraphics[width=3.0in]{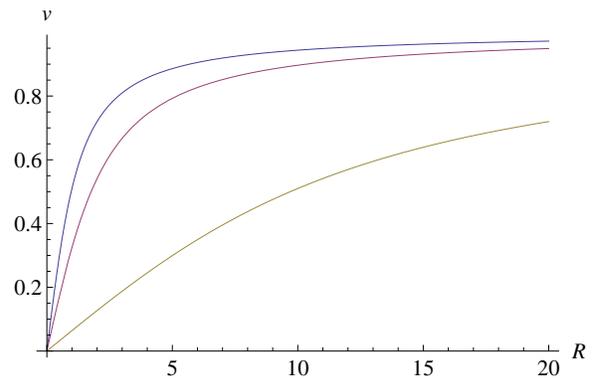}
    \caption{Rotation curves in units with $v_0=1$ in the equatorial
      plane for the missing model (\ref{eq:newdens}) with $b/a = 0.1,
      1$ and 10. As $b/a \rightarrow 0$, the rotation curve attains
      its asymptotic value more quickly.}
    \label{fig:rotfigure}
\end{figure}
\begin{table*}

\null
\vskip\baselineskip
\vskip\baselineskip

$$
\vbox{\tabskip =0pt 
      \offinterlineskip
\def\tablerule{\noalign{\hrule}}
\halign to 430pt{\strut#& \vrule#\tabskip = 1em plus 2em&
\hfil#\hfil& \vrule \hfil # \hfil & \hfil
#\hfil& \vrule#\tabskip=0pt\cr\tablerule
&&\multispan3 {} &\cr
&&\multispan3\hfil GENEALOGY\hfil&\cr
&&\multispan3{} &\cr \tablerule
&&{}&&{}&\cr
&&{${\displaystyle {GM\over r}}$}&& $-v_0^2\log r $ &\cr
&&{}&&{}&\cr
&&Newtonian potential&& Isothermal models &\cr
&&{}&&{}&\cr\tablerule
&&{}&&{}&\cr
&&{${\displaystyle {GM\over (r^2 + a^2)^{1/2}}}$}&& $-v_0^2\log\left[a
+ \sqrt{r^2+a^2} \right]$ &\cr
&&{}&&{}&\cr
&&Plummer (1911) model && Cored Isothermal models &\cr
&&{}&&{}&\cr\tablerule
&&{}&&{}&\cr
&&{${\displaystyle {GM\over r + a}}$}&& $-v_0^2\log\left[a
+ r \right]$ &\cr
&&{}&&{}&\cr
&&Hernquist (1990) model && Evans \& Williams (2014) &\cr
&&{}&&{}&\cr\tablerule
&&{}&&{}&\cr 
&&$ {\displaystyle {GM \over [R^2+(a+|z|)^2]^{1/2}}  }$&& $-v_0^2\log \left[
a +|z| + \sqrt{R^2 + (a+|z|)^2}\right]$ &\cr
&&{}&&{}&\cr
&&Kuzmin (1956) model&& Mestel (1963) model 
&\cr
&&{}&&{}&\cr\tablerule
&&{}&&{}&\cr 
&&${\displaystyle
 {GM \over (R^2 + (a + Z )^2)^{1/2}}},\qquad Z^2 = b^2 +z^2$&&${\displaystyle 
 -v_0^2 \log \left[ a + Z + \sqrt{ R^2 + (a + Z)^2} \right]}.
$&\cr
&&{}&&{}&\cr
&&Miyamoto Nagai (1975) model && This paper &\cr
&&{}&&{}&\cr\tablerule
\hfil\cr}}
$$


    \caption{The genealogy of models; the models in the right column
      are derived from the ones in the left by integration with
      respect to the model parameter $a$, together with the
      identification $v_0^2 = GM/a$. The Plummer (1911) model sires
      the family of cored isothermal potentials~\citep{Wi99}. In the
      limit $a\rightarrow 0$, the reduces to the Newtonian potential
      siring the singular isothermal sphere. The \citet{He90}
      model generates the cusped halo family recently discussed by
      \citet{Ev14}. Likewise, the Kuzmin (1956) disc yields the family
      of cored Mestel discs (Evans \& Collett 1993), which become
      Mestel's (1963) model as $a\rightarrow 0$. Finally, the Miyamoto
      \& Nagai (1975) disc sires the flat rotation curve models
      discussed in this paper.}
\label{tab:familytree}
\end{table*}

\section{Extremely Flat Halo Models}

\subsection{A Little History}

\citet{Mi75} and \citet{Na76} devised an ingenious way to build simple
models of flattened galaxies. The models have become widely
used. There are two reasons for this. First, the potential, and even
more importantly, the force components are derived from simple,
analytic functions. So they can speed N-body simulations with economy
of effort. Second, the models may become arbitrarily flattened and so
can represent the gamut of shapes of galaxies from spherical to
disc-like.

Miyamoto \& Nagai's (1975) starting point was the desire to
thicken up the infinitesimally thin \citet{To63} discs. This family
of discs has density
\begin{equation}
\label{eq:toomre}
\rho_{\rm T}(R,z) = \Sigma(R)\delta(z) = {\Sigma_0 a^{2n +1} \over (R^2
  +a^2)^{n+\half}}\delta(z),\qquad n>0.
\end{equation}
When $n$ is an integer greater than zero, Toomre (1963) showed that
the corresponding potential is analytic. The simplest of the Toomre
discs, corresponding to $n=1$, was previously discovered by Kuzmin
(1956) and has the potential.
\begin{equation}
\label{eq:kuzmin}
\psi_{\rm T}(R,z) = {GM \over (R^2 + (a + |z|)^2)^\half}.
\end{equation}
The potentials of the higher Toomre discs can all be derived from the
Kuzmin disc by differentiation with respect to the parameter
$a^2$~\citep[see e.g.][]{BT}.

Miyamoto \& Nagai (1975) hit upon the idea of making the replacement
\begin{equation}
\label{eq:trans}
|z| \rightarrow \sqrt{b^2 + z^2} \doteq Z
\end{equation} 
in the potential of the Toomre discs. This indeed thickens them up and
makes them versatile models of flattened galaxies. In particular, the
Kuzmin disc (\ref{eq:kuzmin}), under the transformation
(\ref{eq:trans}), becomes exactly the simplest and most widely-used
Miyamoto \& Nagai (1975) model.  In cylindrical polar coordinates ($R,
\phi,z$), it has a gravitational potential
\begin{equation}
    \label{eq:pot}
    \psi_1(R,z) = {GM \over (R^2 + (a + Z)^2 )^{1/2}}, \qquad\qquad Z^2 = b^2
    + z^2,
\end{equation}
where $M$ is the mass of the galaxy, and $a$ and $b$ are two
scalelengths. Using Poisson's equation, the density corresponding to
the potential (\ref{eq:pot})
\begin{equation}
    \label{eq:dens}
    \rho_1(R,z) = {b^2M\over 4 \pi} {aR^2 + (a+ 3Z)(a+Z)^2 \over Z^3(R^2
      + (a+Z)^2)^{5/2}}.
\end{equation}
Asymptotically, the density falls off like $\rho \sim R^{-3}$ (except
if $a=0$) along the equatorial plane and like $\rho \sim z^{-5}$ along
the pole.  When $a=0$, the potential reduces to that of a
spherical~\citet{Pl11} model.  When $b=0$, the potential is that of a
razor-thin \citet{Ku56} disc. The potential (\ref{eq:pot}) therefore
provides an entire family of models which vary continuously between
the Plummer sphere and the infinitesimally thin Kuzmin disc.

Just as the $n$th Toomre (1963) disc can be generated by
differentiation $n$ times with respect to $a^2$, so \citet{Na76}
demonstrated that there is an entire family of thick disc models that
can be generated from (\ref{eq:dens}) by the same procedure.  The
family members satisfy the recurrence relations~\citep{Na76}
\begin{eqnarray}
\psi_{n+1}(R,z) &=& \psi_{n} - {a\over 2n-1} {\partial \over \partial
  a}\psi_{n},\label{eq:recura}\\
\rho_{n+1}(R,z) &=& \rho_{n} - {a\over 2n-1} {\partial \over \partial
  a}\rho_{n}.
\label{eq:recurb}
\end{eqnarray}
They are known collectively as the Miyamoto \& Nagai models~\citep[see
  e.g.][]{BT}. The $n$th member of the family has density falling off
like $R^{-2n-1}$ in the equatorial plane.

We shall now show that there is a model that has been missing all
these years. It is the topmost model of the Miyamoto \& Nagai family,
corresponding to $n=0$, and it has a flat rotation curve.

\subsection{The Missing Model}

\citet{Ev92} discovered a simple potential-density pair for a disc
with an asymptotically flat rotation curve
\begin{equation}\label{eq:rybicki}
\Sigma (R) = {\Sigma_0 a\over \sqrt{a^2 + R^2} },\qquad
\psi(R,z) = -v_0^2 \log \left[ a + |z| + \sqrt{ (a + |z|)^2 + R^2} \right],
\end{equation}
where $v_0^2 = 2\pi G \Sigma_0 a$~\citep[see also][]{Ev93}. In the
limit $a \rightarrow 0$ while maintaining $v_0$ fixed, the \citet{Me63}
scale--free disc
\begin{equation}
\label{eq:mestel}
\Sigma(R) = {\Sigma_0 a \over R},\qquad \psi(R) = -v_0^2 \log (R),
\end{equation}
is recovered. As Evans \& de Zeeuw (1992) pointed out, the model
(\ref{eq:rybicki}) can be considered as a missing Toomre disc. It is
really the $n=0$ member of the family (\ref{eq:toomre}). No doubt
because it has the defect of infinite mass, it was not considered as
an interesting model in the era before the discovery of flat rotation
curves.

Proceeding analogously to Miyamoto \& Nagai (1975), we thicken up the
missing Toomre disc and arrive at the potential
\begin{equation}
\label{eq:newpot}
\psi_0(R,z) = -v_0^2 \log \left[ a + Z + \sqrt{ (a + Z)^2 + R^2} \right],\qquad Z^2 = b^2 + z^2.
\end{equation}
Application of Poisson's equation gives the three dimensional density as
\begin{equation}
\label{eq:newdens}
\rho_0(R,z) = {v_0^2 b^2 \over 4 \pi G}{R^2 + (a+2Z)(a+Z)\over
  Z^3(R^2 + (a+Z)^2)^\threeh}.
\end{equation}
Here, $b>0$ without loss of generality. Then, the density is
everywhere positive definite provided $a+b>0$.  It is this model that
we shall study in the remainder of the paper. It is the missing
Miyamoto \& Nagai model.

To substantiate our claim that this model should truly be considered
part of the Miyamoto \& Nagai (1975) family, we need only recall that
each succeeding member of the sequence is generated by differentiation
with respect to the scalelength $a$. So, if our model is indeed a
missing relative, we expect it to satisfy the recurrence relations
~(\ref{eq:recura}) and ~(\ref{eq:recurb}) with $n=0$. It can be easily
verified that on setting $v_0^2 = GM/a$ such is indeed the case.

Table~\ref{tab:familytree} summarises the genealogy of some of the
models discusses in this Section. The device of integration or
differentiation with respect to the scalelength $a$ is simple, but
surprisingly powerful.  All the known flat rotation curve models can
be derived from simple variants of the point mass potential (such as
Plummer or Kuzmin) by this means.

Finally, we note that \citet{Zo11} has introduced a related model
described as ``a combination of the logarithmic potential and a
Miyamoto-Nagai model''. It has potential
\begin{equation}
\psi(R,z) = -{v_0^2\over 2} \log \left[ (a + Z)^2 + R^2\right],\qquad Z^2 = b^2 + z^2.
\end{equation}
In the limit $b \rightarrow 0$, this does not correspond to a solution
of Laplace's equation everywhere except $z=0$, so it does not become
an infinitesimally thin disc. Therefore, the model has properties that
are rather different from the sequences studied by \citet{Mi75} and
\citet{Na76}.

\section{Model Properties}

Here, we describe the properties of the potential-density pair give in
eqns~(\ref{eq:newpot}) and (\ref{eq:newdens}). Henceforth, to reduce
notational clutter, we drop the subscript zero, which was introduced to
establish that our models are the zeroth member of the Miyamoto \&
Nagai (1975) family.

\subsection{Density}

The density (\ref{eq:newdens}) falls off asymptotically like $\rho \sim
R^{-1}$ in the equatorial plane, and like $\rho \sim z^{-4}$ along the
pole. The models are always highly flattened. The limit $b \rightarrow
0$ gives an infinitesimally thin disc, which becomes the Mestel disc
when additionally $a \rightarrow 0$.

More generally, Taylor expanding (\ref{eq:newdens}) about the origin
tells us that the density there looks like
\begin{equation}
\rho(R,z) = \rho_0\Biggl[ 1 - {R^2\over a_1^2} - {z^2\over a_2^2} -
  \cdots \Biggr],
\end{equation}
with the axis ratio
\begin{equation}
\left({a_2 \over a_1}\right)_0^2 = {b^2 (a+4b)\over (a+b)(3a^2 + 9ab +
  8b^2)}.
\end{equation}
The model with $a = a_{\rm crit} = -\frac{1}{3}(19^{1\over 3} -4) =
-.44387$ has spherical equidensity contours at the centre. If $a >
a_{\rm crit}$, the models are oblate. If $-b<a< a_{\rm crit}$, the
models are prolate at the centre.  Irrespective of the behavior at the
centre, the models all become extremely flattened and highly oblate,
with an axis ratio $a_2/a_1 \rightarrow 0$ at large radii. In other
words, this family can exhibit a variety of behaviour near the centre,
but is always disc-like at large radii.

Contours of constant density for some of the models are shown in
Fig.~\ref{fig:densfigure}. Note that they exhibit a range of behaviour
including extreme flattenings. For modelling of galaxies, we remark
that this represents the density contours of both the luminous and
dark components, as the model already has a flat rotation curve. It is
therefore straightforward to carve out a luminous exponential disc if
we wish to separately identify and study the individual components.

\begin{figure}
    \centering \includegraphics[width=2.5in]{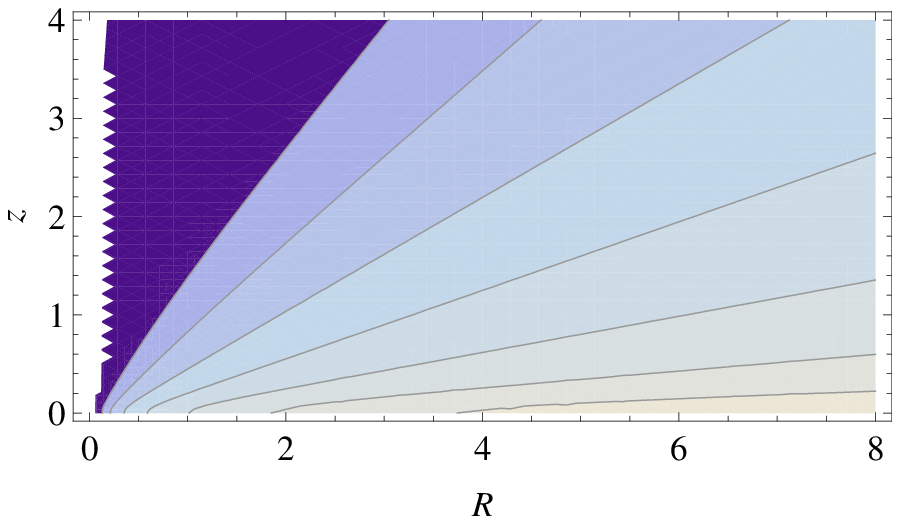}
    \includegraphics[width=2.5in]{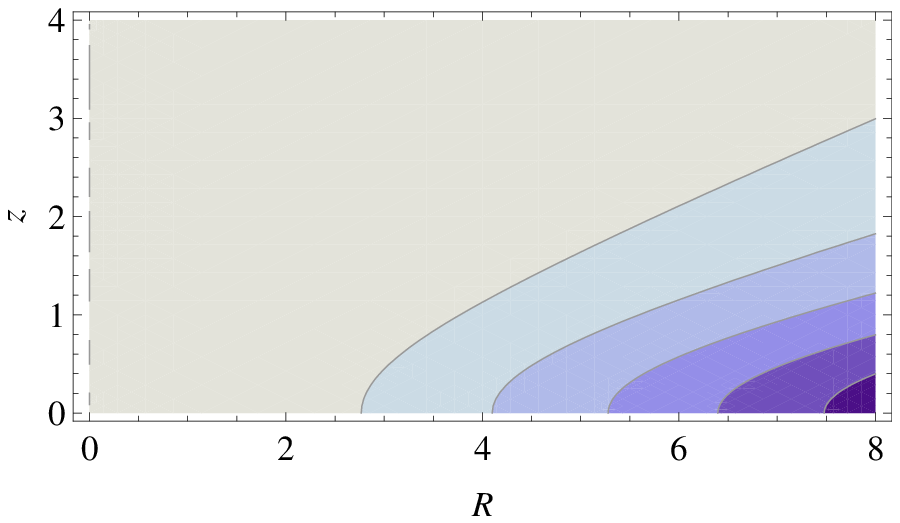}
    \includegraphics[width=2.5in]{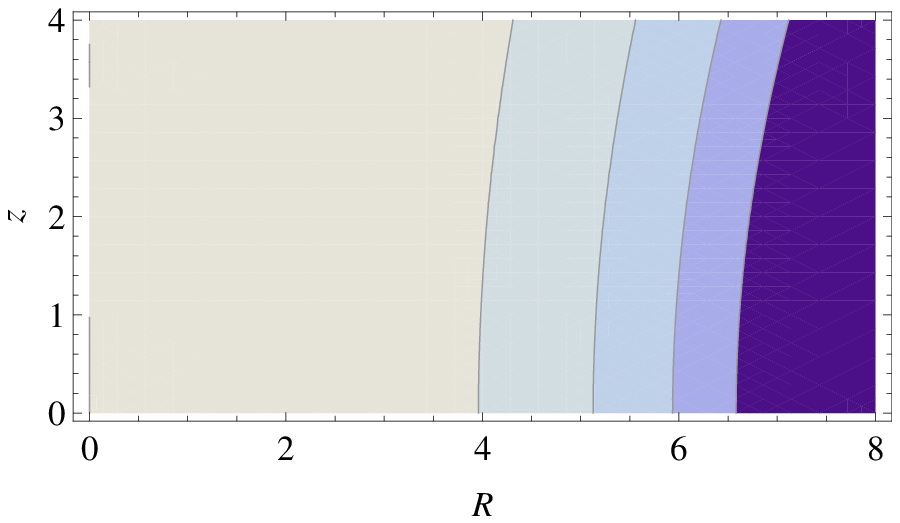}
    \caption{Contours of anisotropy parameter $\beta$ in ($R,z$) plane
      for the model (\ref{eq:newdens}) with $b/a = 0.1, 1$ and
      10. With increasing flattening, the models become strongly
      tangentially anisotropic. In the uppermost panel, contours run
      from -0.1 to -1000, the anisotropy parameter increasing by a
      factor of 10 from one contour to the next. The model is dominate
      by extreme tangential anisotropy.  In the middle panel, contours
      run from 0 to -10 in steps of -2, so the tangential anisotropy
      is significant but not overwhelming.  In the lower panel,
      contours run from 0 to -0.08 in steps of -0,02, so the model is
      still nearly isotropic.}
    \label{fig:aniso}
\end{figure}

\subsection{Kinematics}

The force components are
\begin{eqnarray}
F_R &=& - {Rv_0^2\over \sqrt{R^2 + (a+Z)^2} (a+Z + \sqrt{R^2
    +(a+Z)^2})},\nonumber\\
F_z &=& - {zv_0^2\over Z \sqrt{R^2 + (a+Z)^2}}.
\end{eqnarray}
The rotation curve in the equatorial plane is
\begin{equation}
v_{\rm circ}^2(R) = v_0^2 \left[ 1- {A \over
  \sqrt{R^2 + A^2}}\right], \qquad A = a+b.
\label{eq:rotcurve}
\end{equation}
All the models have asymptotically flat rotation curves $v_{\rm circ}
\rightarrow v_0$. Just as for the Miyamoto \& Nagai models, the
rotation curves are identical for all models with the same $A=a+b$. In
particular, if $a+b=0$, the rotation curve is everywhere constant and
equal to $v_0$.  Some sample rotation curves are illustrated in
Fig.~\ref{fig:rotfigure}.

By Jeans theorem~\citep[see e.g.,][]{BT}, the distribution function
$F$ can only depend on the integrals of motion. For this axisymmetric
potential, the two classical integrals are the binding energy and the
angular momentum per unit mass.  The velocity second moments for the
two-integral $F(E,L_z)$ can be found analytically, either by solving
the Jeans equations or by using the \citet{Hu77} formulae, which are
\begin{eqnarray}
\rho\langle v_R^2 \rangle &=& \rho \langle v_z^2 \rangle = \int_{-\infty}^\psi \rho(\psi',R^2)
d\psi',\\
\rho\langle v_\phi^2 \rangle &= & \int_{-\infty}^\psi {\partial \over
  \partial R}\bigl( R \rho(\psi',R^2)\bigr)
d\psi',
\end{eqnarray}
Here, the density has been written as an explicit function of $R$ and
$\psi$, rather than $R$ and $z$. This can be done, as
\begin{equation}
Z(z) = \half \left( \exp(-\psi/v_0^2) - R^2\exp(\psi/v_0^2) \right) -a,
\end{equation}
Substituting into (\ref{eq:newdens}) gives $\rho(\psi,R^2)$. The
integrations in Hunter's formulae can be explicitly performed to give
an analytic solution of the Jeans equations as
\begin{equation}
\langle v_R^2 \rangle = \langle v_z^2 \rangle = {v_0^2\over 2}\,
        {Z\sqrt{R^2 + (a+Z)^2} \over R^2 + (a+Z)(a+2Z)}
\end{equation}
\begin{equation}
\langle v_\phi^2 \rangle = {v_0^2\over 2}\, {2R^2 + 2(a+Z)(a+2Z) -(2a+3Z)\sqrt{R^2 + (a+Z)^2} \over
  R^2 + (a+Z)(a+2Z)}
\end{equation}
The velocity second moment tensor is aligned in cylindrical polar
coordinates so the cross-terms vanish, i.e., $\langle v_R v_z\rangle=
0 = \langle v_Rv_\phi \rangle = \langle v_zv_\phi\rangle$.  The models
are isotropic on the pole $R=0$, but everywhere else, $\langle
v_\phi^2 \rangle > \langle v_R^2 \rangle$. In the equatorial plane as
$r\rightarrow \infty$, $\langle v_R^2\rangle \rightarrow0$ whilst
$\langle v_\phi^2 \rangle \rightarrow v_0^2$.  Contours of the second
moment anisotropy parameter
\begin{equation}
\beta = 1 - {\langle v_\phi^2 \rangle \over \langle v_R^2 \rangle}
\end{equation}
are shown in Fig.~\ref{fig:aniso}. In the equatorial plane, as $R
\rightarrow \infty$, $\beta \rightarrow -2R/b$ and so the model
becomes increasingly dominated by circular orbits.

\begin{figure}
    \centering
    \includegraphics[width=2.5in]{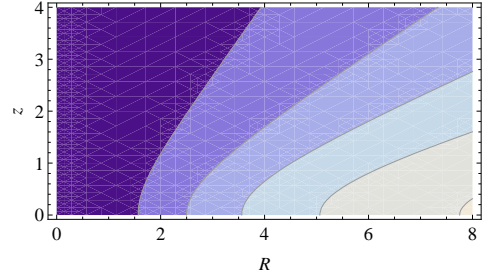}
    \caption{Contours of maximum mean streaming velocity in ($R,z$)
      plane for the model (\ref{eq:newdens}) with $b/a = 1$. Contours
      run from 0.4 to 0.8 in units with $v_0=1$.}
    \label{fig:meanv}
\end{figure}

So far, the models have no stellar rotation. The maximum streaming
model is created by reversing all stars with $L_z<0$, so that all
stars rotate in the same direction.  The maximum mean streaming
velocity at any point ($R,z$) is
\begin{equation}
\langle v_\phi \rangle = {\sqrt{2}\over \pi \rho} \int_{-\infty}^\psi
{d\psi'\over \sqrt{\psi-\psi'}} \int_0^R {dR'\over \sqrt{R^2-R'^2}}
  {\partial \over \partial R} R\rho(\psi',R'^2)
\end{equation}
This is easily evaluated numerically -- Fig.~\ref{fig:meanv} shows the
maximum streaming velocity for the model with $b/a =1$.

To summarise, the second moments of the model are all analytic, whilst
the maximum mean streaming velocity is a simple quadrature. This
enables models to be built with a wide range of kinematic properties.

\begin{figure*}
    \centering
    \includegraphics[height=3in]{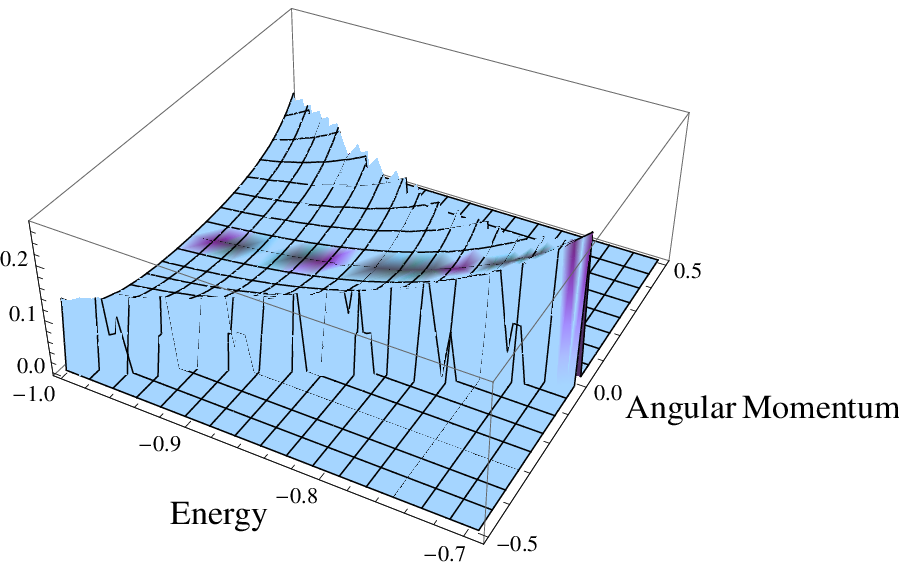}
    \includegraphics[height=3in]{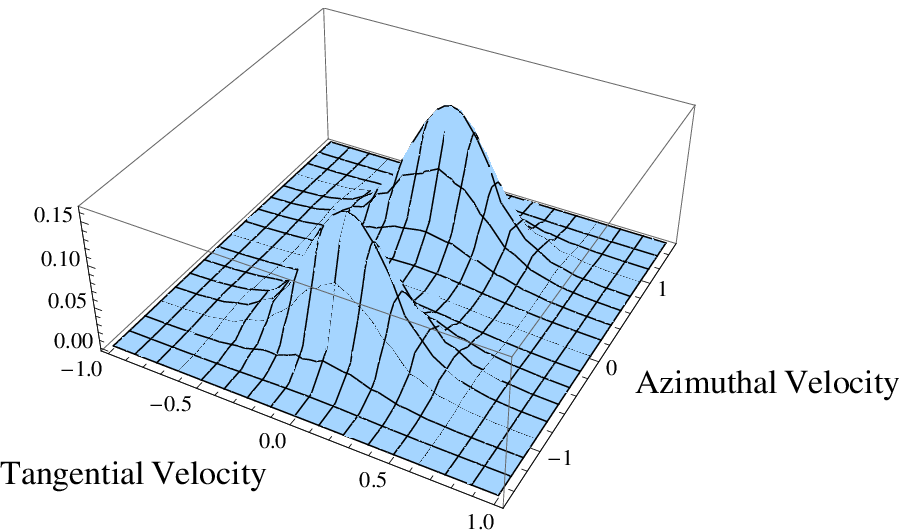}
    \caption{Plots of the even part of the distribution function for
      the model (\ref{eq:newdens}) with $b/a= 1$. Upper panel: the DF
      in integral space ($E,L_z$). Lower panel: the velocity
      distribution ($v_\phi, v_{\rm T}$) in the equatorial plane at
      $(R,z) = (2a,0)$. Here, $v_{\rm T} = \sqrt{v_R^2 +v_z^2}$.}
    \label{fig:evendf}
\end{figure*}
\begin{figure*}
    \centering
    \includegraphics[width=3.5in]{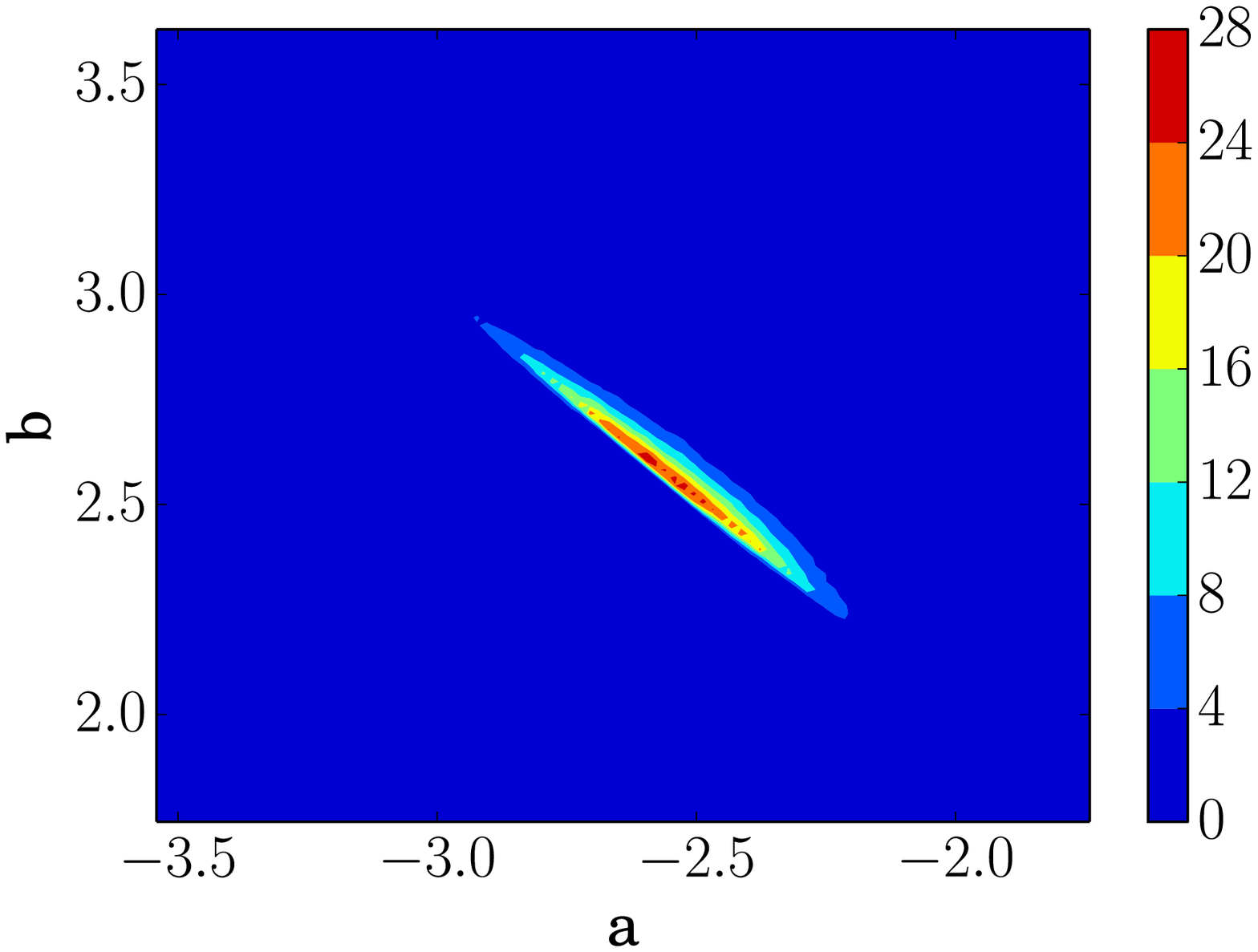}
    \includegraphics[width=3.5in]{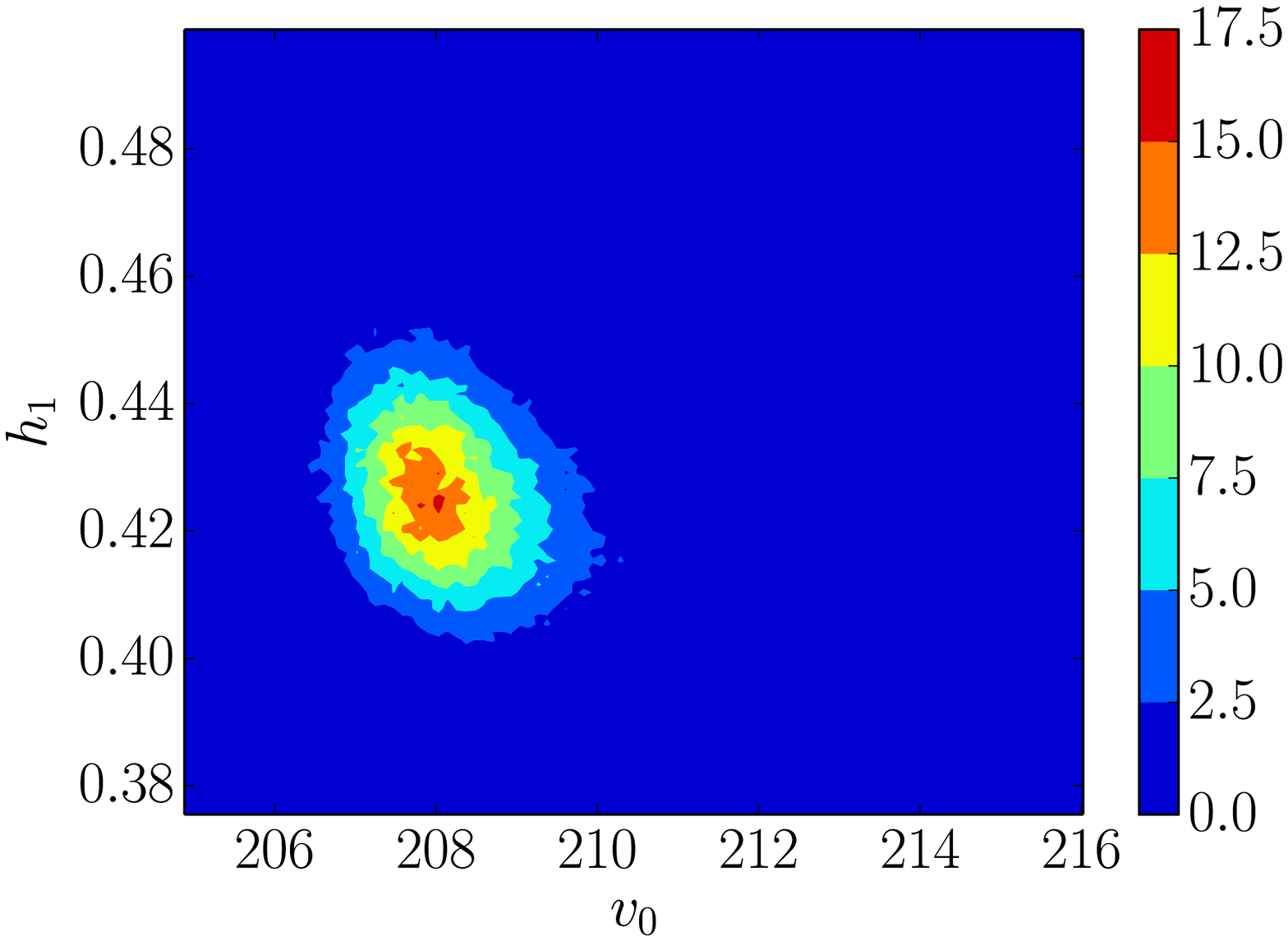}
    \includegraphics[width=3.5in]{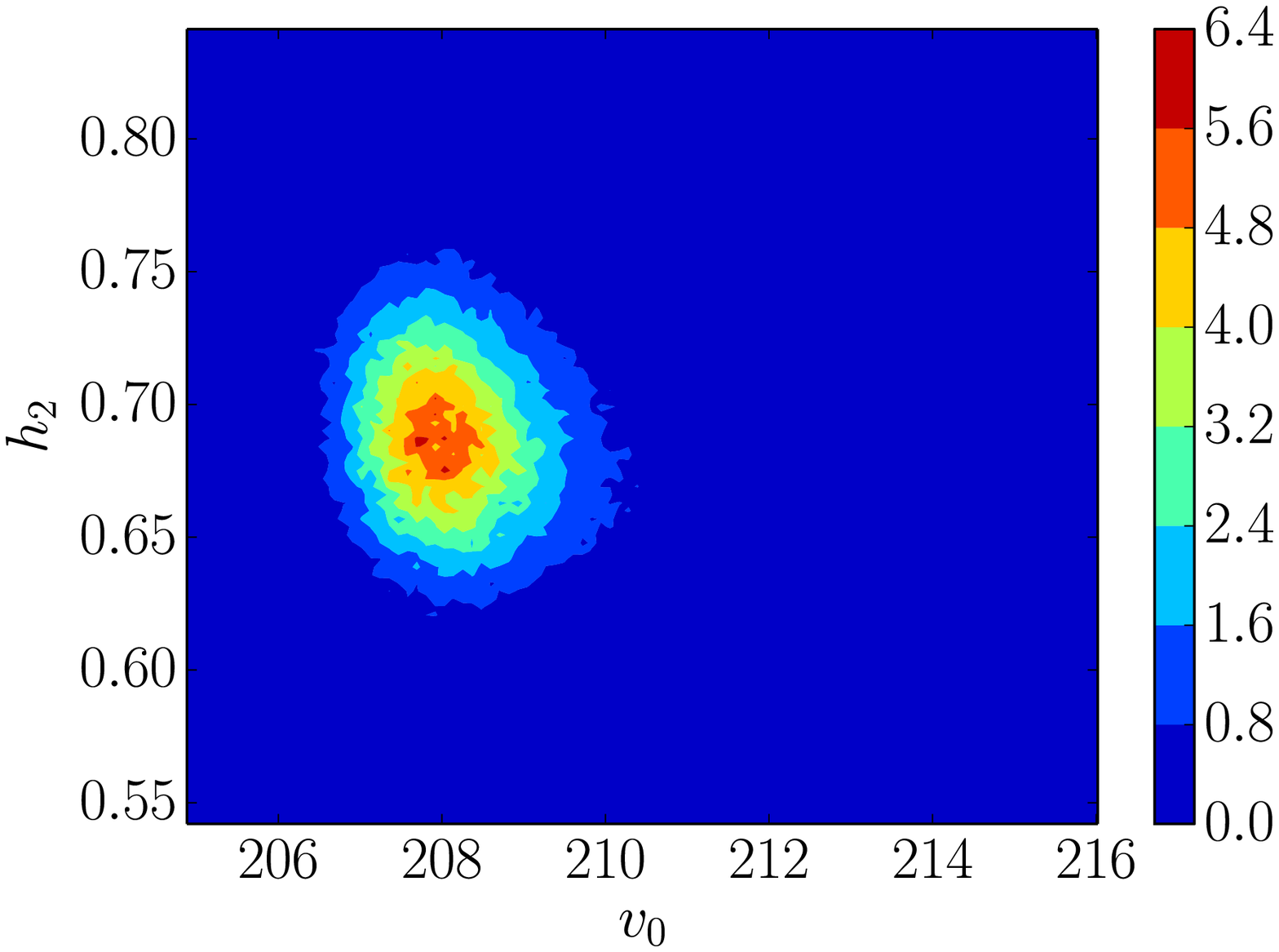}
    \caption{The model (\ref{eq:newdens}) is fitted to the rotation
      curve data of \citet{So09} and the vertical velocity dispersion
      data of populations of thick disc stars given in \citet{Sm12}
      and \citet{Mo12}. The plots show likelihood contours in the
      plane of the recovered model parameters $v_0$, $a$ and $b$,
      together with the scale heights of the two populations $h_1$ and
      $h_2$. The colour scheme shows the normalized posterior
      probability density.
      The posterior values for the model parameters are $v_0 =
      208.3^{+0.7}_{-1.2}$ kms$^{-1}$, $a = -2.53^{+0.20}_{-0.18}$ and
      $b = 2.58^{+0.17}_{-0.20}$, whilst the scaleheights are $h_1 =
      426^{+12}_{-14}$ pc and $h_2 = 689^{+31}_{-37}$ pc
      respectively.}
    \label{fig:params}
\end{figure*}

\begin{figure}
     \centering
    \includegraphics[width=2.5in]{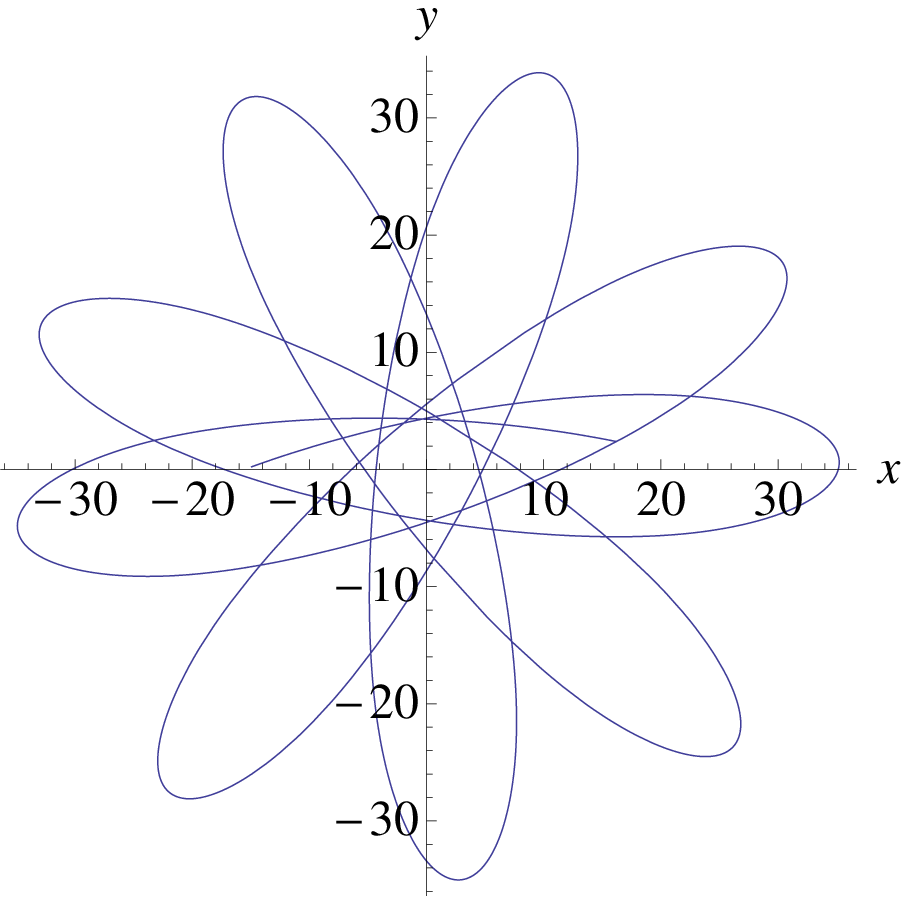}
    \includegraphics[width=2.5in]{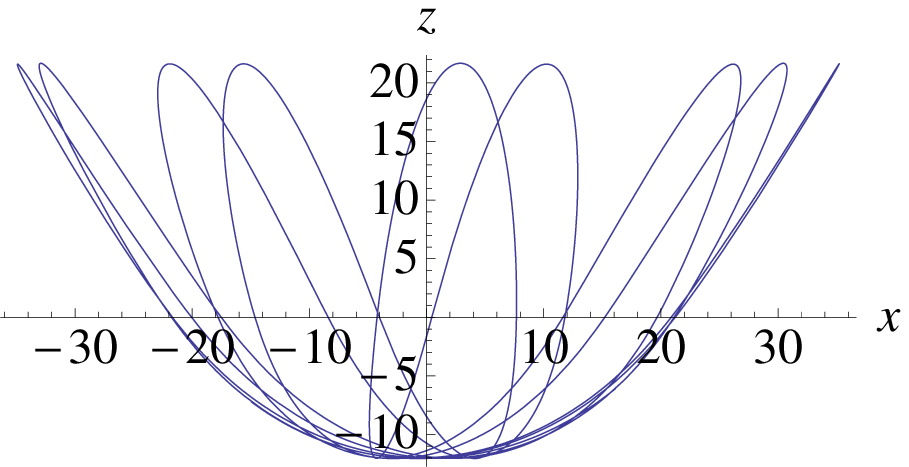}
    \includegraphics[width=2.5in]{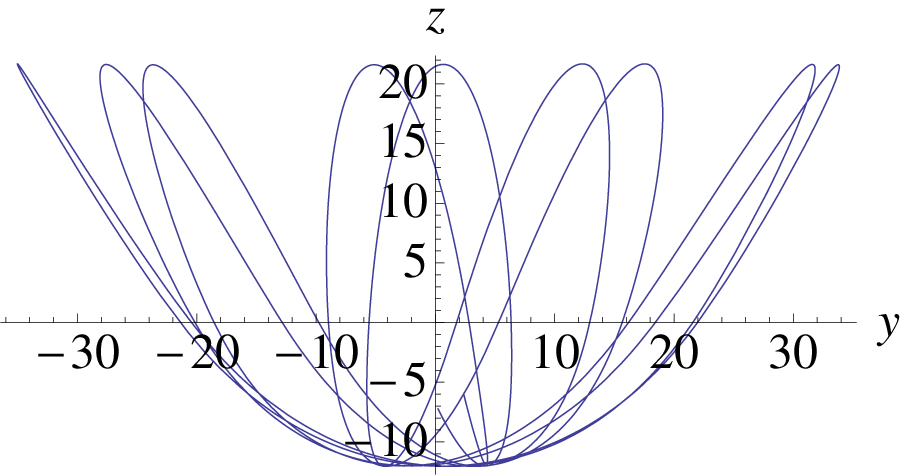}
\vskip7pt
    \includegraphics[width=2.5in]{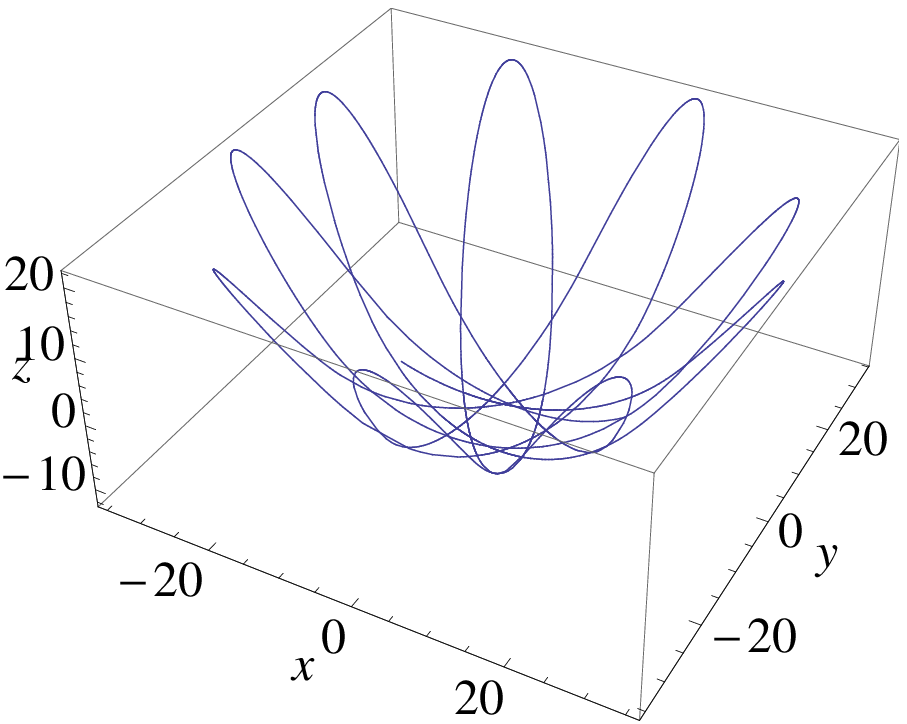}
    \caption{Plots showing the orbit of the Sgr for the best fit model
      parameters in the three principal planes. The orbit has a
      pericentric distance of 0.6 kpc, apocentre of 60.2 kpc and a
      maximum height above the disc of 30.8 kpc. The integration time
      is $\sim 5$ Gyr. The trajectory is recognized as a
      precessing banana orbit.}
    \label{fig:sagorbit}
\end{figure}

\subsection{The Distribution Function}

For some applications -- such as studies of the stability of highly
flattened galaxies ~\citep{Me94} -- it is useful to have the full
distribution function (DF) of the model. For our application, we do
not need the precise form of the DF, but we do need to be certain that
the models are physical and correspond to an everywhere positive DF.
This is not a matter of idle pettifoggery ! The well-known flattened
logarithmic model fails to be physical once its equipotential axis
ratio is less than $1/\sqrt{2} \approx 0.707$, and so cannot attain
very extreme ellipticities~\citep{Eva93, BT}. Without constructing
the DF and checking it is positive, we cannot be sure that our
potential-density pair really do correspond to viable halo models.

For our models, the density $\rho(\psi,R^2)$ is
\begin{eqnarray}
\rho(\psi,R^2) &=& {4b^2 v_0^2\over \pi
  G}\Biggl[{\exp{4\psi}\over
(1+R^2\exp(2\psi))(1-2a\exp(\psi)-R^2\exp(2\psi))^3} \Biggr.\nonumber \\
&+&\Biggl.
{\exp{4\psi}\over
(1+R^2\exp(2\psi))^3(1-2a\exp(\psi)-R^2\exp(2\psi))^2}\Biggr]
\label{eq:densrho}
\end{eqnarray}
The even part of the two-integral DF is given by \citet{Hu93} as the
contour integral
\begin{equation}
\Fe(E,L_z^2) = {1\over 4\pi^2 i \sqrt{2}} {\partial^2 \over \partial E^2}
\int_{-\infty}^{E+} {d\psi \over (\psi-E)^{1/2}} \rho\Biggl(\psi,
  {L_z^2\over 2(\psi-E)}\Biggr).
\end{equation}
A cut is needed in the complex plane from $\psi =E$ to $\psi =
-\infty$ to define the square root function. The contour is a loop
that starts on the lower side of the real $\psi$ axis at $\psi =
-\infty$. The circle encircles $\psi = E$ in the positive sense,
crossing the real axis at $\psi = \psie(E)$ and ending at
$\psi = -\infty$ on the upper side of the real axis. Here,
\begin{equation}
\psie(E) = \psi(\Rc(E),0)
\end{equation}
where $\Rc$ is the radius of the circular orbit in the equatorial
plane with energy $E$. The circular orbits are defined parametrically
by
\begin{eqnarray}
E &=&  -v_0^2\log (A + \sqrt{\Rc^2+A^2}) -
{L_z^2\over 2\Rc^2} \nonumber \\
L_z^2 &=& {\Rc^4v_0^2 \over \sqrt{\Rc^2 + A^2} (A + \sqrt{\Rc^2 + A^2})}
\end{eqnarray}
So, $\psi_{\rm env}$ is the root of the equation
\begin{equation}
2(\psie -E) - v_0^2 + {Av_0^2 \over \exp(-\psie/v_0^2) - A} =0 
\end{equation}

As the density in eq.~(\ref{eq:densrho}) is a rational function, there
are no further branch cuts needed and so it is straightforward to
devise a contour. For ease of numerical evaluation, it must not come
too close to the cut along the real axis. For our calculations, we
choose to parametrize the bottom half of the contour by
\begin{equation}
\psi = y + \psi_{\rm env} (E) + i \epsilon (\exp (y) -1),\qquad\qquad
-\infty \le y \le 0,
\end{equation}
where $\epsilon$ is small. We then obtain the distribution function by
doubling the real part along it.

The DF is shown in Fig.~\ref{fig:evendf} both in the space of
integrals ($E,L_z$) and in velocities ($v_\phi, v_{\rm T}$), where
$v_{\rm T} = \sqrt{v_R^2+v_z^2}$). The flattening of the model at
large radii is produced by the bias towards circular orbits, as
indicated by the rise of the DF towards the edge of the domain in the
upper panel. As this is an even DF, there are equal numbers of
rotating and counter-rotating stars, and these show up as two peaks in
the velocity distributions in the lower two panels corresponding to
near-circular orbits. It is of course straightforward to produce discs
of stars rotating in one direction, for example, by doubling the part
of the DF corresponding to positive $v_\phi$ and nulling the part
corresponding to negative $v_\phi$. 

Evaluations on a numerical grid confirm that the even DF is everywhere
positive. This guarantees that the underlying model is physical, even
for very severely flattened configurations.

\section{Application: The Flattening of the Galaxy}

\subsection{Data}

Two independent constraints on the potential of the Galaxy are the
rotation of the HI gas in the Galactic plane and the kinematics of
stars perpendicular to the plane.  These data have been examined many
times before. Most previous investigations have used a model that is
decomposed into disc and spherical halo with the aim of constraining
the local dark matter density.  Here, our focus is different. We will
not make any distinction between baryonic and dark matter and seek to
put constraints on the overall flattening of the potential and density
of the Galaxy.

The study of the kinematics of vertical tracers of stars near the Sun
goes back to the very origins of the dark matter problem~\citep{Oo32}.
The underlying idea is to use a volume complete survey of stars with
known distances and kinematics to constrain the gravitational
potential either with distribution functions or with the Jeans
equations~\citep[see e.g.,][]{Ku89b,Ku89a,Ho04,Sm12}. Often, the
potential is parametrised in a simple way, encoding prior beliefs as
to the likely density of disc and halo. So, for example, \cite{Ku89a}
use a simple potential mimicking the effects of disc and halo,
together with a prior on the rotation curve\footnote{The
  model~(\ref{eq:newpot}) has exactly the form proposed in eq.(40) of
  \citet{Ku89a} at small heights from the Galactic Plane}.

By itself, the rotation curve of the Milky Way provides no constraint
on the flattening of the density, as models that are spherical,
spheroidal or disc-like can all possess flat rotation curves. From the
perspective of the local dark matter density, this has been recently
emphasized by \citet{We10}. They use a variety of spherical and
spheroidal halo models coupled with baryonic thin and thick
exponential discs to reproduce the rotation curve and find the local
dark matter density is uncertain by at least a factor of 2. Here, we
make no attempt at a decomposition into disc and halo components, but
model the entire potential and density of the Galaxy by
eqs~(\ref{eq:newpot}) and (\ref{eq:newdens}). Our aim is to understand
how flattened the density contours of the Galaxy could possibly be,
given the constraints from the vertical kinematics of stars in the
solar neighbourhood and the behaviour of the rotation curve from 4kpc
to 20 kpc.

We use the data on the vertical kinematics from two sources. First,
\citet{Sm12} use samples of dwarf stars extracted from the Sloan
Digital Sky Survey Stripe 82 catalogue (Bramich et al. 2008). They
provide the vertical and radial velocity dispersions as a function of
$z$ up to 2.5 kpc below the Galactic plane.  Second, \citet{Mo12}
provide the variation of velocity dispersion with $z$ using the
velocities of $\sim 400$ red giants towards the South Galactic
Pole. These thick disc stars extend to heights of up to 4.5 kpc from
the Galactic plane.

Complementing this, we also require that the models match the rotation
curve of the Galaxy. For the inner Galaxy, this is determined by the
terminal velocities of HI and CO gas , for the outer Galaxy from
spectrophotometry of HII and CO regions ~\citep[e.g.,][]{Fi89}. The
variety of methods does lead to some scatter, but \citet{So09} has
recently compiled and re-analysed the data, calibrating it to Galactic
constants of Solar position of 8 kpc and Local Standard of Rest of 200
kms$^{-1}$.

\begin{figure}
     \centering
    \includegraphics[width=3.in]{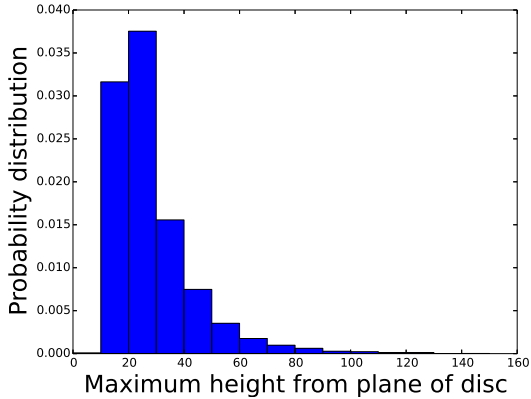}
  \caption{Histogram showing the maximum distance from the plane of
    the disc reached by $10^4$ Sgr orbits, integrated backwards with
    starting condition drawn from the current error circle on position
    and velocity. Only 4.3 \% of the orbits reach a height greater
    than 50 kpc, and 2.25 \% a height greater than 60 kpc.}
\label{fig:zmax}
\end{figure}

\subsection{Results}

Given the model parameters $v_0, a, b$, the rotation curve in the
Galactic plane ($z=0$) is given by eqn~(\ref{eq:rotcurve}). We can
also calculate the vertical velocity dispersion for a tracer
population as a function of $z$ at the radius $R=R_{\odot}$ via direct
integration of the vertical Jeans equation, assuming the vertical
density distribution of the tracers is known. We have two sources of
data for the vertical velocity dispersion~\citep{Sm12,Mo12} and we
assume these data are well represented by two distinct exponential
density distributions with scale heights $h_1$ and $h_2$. This is a
common assumption, supported by the original work identifying the
Galactic thick disc ~\citep{Gi83}, as well as photometry of external
thick discs in edge-on galaxies, which show exponentially declining
profiles at large heights~\citep{Yo06}. Nonetheless, hard evidence
that the density law of thick disc populations in our Galaxy declines
exponentially at these heights is actually rather elusive.

Together with the 3 model parameters, the two disc scale-heights
provide a 5-dimensional parameter space.  The calculated values of the
rotation curve and vertical velocity dispersion can then be compared to
our data, with the likelihood value for these parameters given by
$\exp(-\chi^2)$.

In order to explore this parameter space, we use the open source Monte
Carlo Markov Chain code $emcee$ provided by \citet{Fo13}. Contour
plots of the recovered potential and scale-height parameters are shown
in Fig.~\ref{fig:params}. The posterior values for the model
parameters, together with the $1\sigma$ bounds, are $v_0 =
208.3^{+0.7}_{-1.2}$ kms$^{-1}$, $a = -2.53^{+0.20}_{-0.18}$, $b =
2.58^{+0.17}_{-0.20}$. The scaleheights of the thick disc stars using
the data of \citet{Sm12} and \citet{Mo12} are $h_1 = 426^{+12}_{-14}$
pc and $h_2 = 689^{+31}_{-37}$ pc respectively.  The flatness of the
rotation curve provides the constraints on $v_0$ and $a+b$, whilst the
vertical velocity data constrains the population scale heights and the
individual values of $a$ and $b$. Flatter density distributions than
this can recover the rotation curve, but then they underpredict the
run of vertical velocity dispersion at large heights ($\sim 4$ kpc)
from the Galactic plane.  In our best fit model, the axis ratio of the
equidensity contour that passes through the Solar position is
$0.57$. Models with more extreme flattening than this are ruled out.

To confirm our result that extreme flattenings are not possible, we
can use the Sagittarius (Sgr) tidal stream. In a spherical potential,
the debris from the Sgr stream would be confined to a plane.  In a
nearly-spherical potential, the plane slowly precesses. When the
gravitational field is highly flattened, then there are two
problems. First, the potential may generate more precession than is
actually observed. This idea was first proposed by \citet{Jo05}, who
measured a difference between the mean orbital poles of the leading
and trailing debris of $10.4 \pm 2.6^\circ$ as delineated by M giants
from the Two Micron All-Sky Survey. The much richer and higher quality
Sloan Digital Sky Survey data has recently been re-analysed by
\citet{Be14}, who found a more complex pattern of differential
precession in the bright and faint Sgr streams. For our purposes here,
we summarise Figure 13 of \citet{Be14} by the statement that the
magnitude of the precession of the wraps of the leading and trailing
debris is certainly less than $15^\circ$.  Secondly, the present day
kinetic energy of the Sgr dwarf is known to reasonable accuracy, as
its proper motion has been measured~\citep{Di05}. If the
equipotentials are highly flattened, then the Sgr cannot rise high
enough above the Galactic plane to provide the debris in the tails. We
note that Sgr stream stars have been detected at heights of $\sim 60$
kpc above the Galactic plane by~\citep[see e.g.,][]{Ma03,Be14}. This
is a serious challenge if the halo flattening is substantial.

Using our best fit model parameters, we integrate the orbit of the
Sagittarius dwarf Galaxy backwards from its present day position of
$[x,y,z,v_x,v_y,v_z] = [-16.1,2.4,-6.1,232,-47,190]$ in units of kpc
and kms$^{-1}$~\citep[see e.g.,][]{Gi14}. The results of this
integration are shown in Fig.~\ref{fig:sagorbit}, which show the Sgr
moving on a precessing banana orbit. This is a typical orbit for
starting positions within the observational errors.  There are two
immediate and related problems.  First, the view of the orbit in the
($x,y$) plane makes it clear that the precession angle is too
great. From one apocentre to the next, the orbital plane turns through
$\sim 43^\circ$, which is too large to be consistent with the
data. Second, the maximum height of the orbit above the Galactic plane
is just $\sim 31$ kpc. In flattened potentials, the Sgr does not have
enough kinetic energy to reach high enough above the Galactic plane to
generate all the known debris. To emphasise this point, we show in
Fig.~\ref{fig:zmax} a histogram of the maximum heights above the disc
for a range of potential parameters from our posterior distribution,
drawing Sgr co-ordinates from within the current observational
errors. We find that just 2.25 \% of all the orbits reach a height of
greater than $\sim 60$ kpc. With increasing accuracy of the proper
motions, this result could probably be tightened further. Before then,
improved modelling without the assumption that a stream is an orbit is
really needed (e.g., Gibbons et al. 2014). We plan to return to this
problem in a later publication.

Throughout, we have modelled the entire Galactic potential, making no
distinction between the disc and halo components. It follows that our
constraint that $q \gtrsim 0.57$ applies to the equidensity contours
of the total matter distribution. However, the luminous thin and thick
discs of the Galaxy are known to be highly flattened. For example, the
scalelengths of the thin and thick discs are $2.8\pm 0.3$ kpc and $3.7
\pm 0.5$ kpc respectively, whilst the scaleheights range between 350
pc to 1kpc ~\citep{Oh01}. Rather than indulge in uncertain
decompositions into disc and halo, we merely note that -- if the
overall flattening satisfies $q \gtrsim 0.57$ -- then this must
assuredly be true of the dark halo by itself. Indeed, the dark halo
itself must be considerably rounder than our limiting value of E4 or
$q =0.57$)!

\section{Conclusions}

We have introduced a new family of very highly flattened galaxy
models. Although the density contours can be spherical, oblate or
prolate in the central regions, the strongly disc-like nature of the
models rapidly asserts itself at moderate to large radii. The circular
velocity curve of all the models is asymptotically flat. {\it They are
  the only flat rotation curve models known to us whose properties can
  represent three-dimensional thickened disks as well as haloes.}

The family are related to a number of classical models already in the
literature. In the infinitesimally thin limit, they become cored
Mestel discs~\citep{Me63, Ev93}. So, the family can be thought of as
thickened, fully three-dimensional Mestel discs. They are also part of
the Miyamoto \& Nagai (1975,1976) sequence of flattened galaxy models,
albeit missing from the original works which predated the discovery
and widespread acceptance of flat rotation curves.

The kinematic properties of the model are largely analytic. In
particular, the solutions of the Jeans equations in cylindrical polar
coordinates are straightforward. The two integral distribution
function $F(E,L_z)$ has been derived as a quadrature. It is everywhere
positive definite for arbitrarily severe flattenings. The highly
flattened shape is supported by extreme values of the azimuthal
velocity second moment $\langle v_\phi^2\rangle$, or in physical terms
by tangential anisotropy or by rotation.

There has long been a lacuna of models with simple properties that are
highly flattened.  The logarithmic halo model, for example, ceases to
be physical below a flattening in the equipotentials of $q =
1/\sqrt{2}$~\citep{Eva93,BT}, thus preventing exploration of highly
elongated shapes. So, our new models have a number of applications,
for example in the study of the stability of highly flattened galaxies
to bending modes~\citep{Me94} and instabilities of thickened discs
with counter-rotating stars~\citep{Se94}. They also provide natural
models of dark haloes for theories in which the dark matter is cold
atomic or molecular gas~\citep{Pf94a,Pf94b,Da12} or is very highly
flattened, as in the decaying neutrino hypothesis~\citep{Sc93}. They
can be used to assess the claims of highly flattened haloes in some
nearby galaxies, such as NGC 4244~\citep{Ol96}, NCC 891~\citep{Be97}
and NGC 4650~\citep{Co96}.  Another application is to the dark discs
that can form in $\Lambda$CDM as a consequence of satellite
accretion~\citep{Re08}.

We have used the models to study the overall shape of the
Galaxy. Here, the vertical kinematics of stars in the solar
neighbourhood together with the rotation curve constrain the shape of
the density contours to be rounder than $q\approx 0.6$. If the shape
the Galaxy is more highly flattened than this, it is not possible to
reproduce the gently rising run of velocity dispersion with height for
stars above or below the Galactic plane. Confirmation of this is
provided by the Sgr dwarf galaxy. The present-day phase space
coordinates of the Sgr are known. Given this kinetic energy, then if
the halo is flatter than $q\approx 0.6$, the orbit cannot reach $\sim
60$ kpc above the Galactic plane where Sgr debris is detected. Highly
flattened potentials also generate too much orbital plane precession
to be consistent with the debris.

Given that our constraint is on the overall equidensity contours, and
the luminous baryonic material is already known to be highly
flattened, this means that the shape of the dark matter halo alone at
least for our Galaxy must be considerably rounder than E4 (or $q=
0.57$). This therefore rules out theories in which the dark matter is
highly flattened -- in particular the suggestion that cold neutral or
molecular gas in the disc can generate the flat rotation
curve. 

Of course, many papers have studied the flattening of the halo of the
Milky Way, using techniques such as the stellar kinematics of halo
stars~\citep{Sm09,Lo12}, cold stellar streams~\citep{Ko10} and
the Sagittarius stream~\citep{He04, Fe06}. Given the discrepancies
between results, there are clearly substantial degeneracies and
establishing the precise flattening will require careful fitting to
multiple datasets. Nonetheless, some regions of parameter space can be
robustly excluded and some definite results obtained. The virtue of
the approach in this paper is that we have shown that extremely
flattened halo models (ellipticity greater than 0.43) are inconsistent
with both the stellar kinematics and the Sagittarius stream.

\section*{Acknowledgements}
AB thanks the Science and Technology Facilities Vouncil (STFC) for the
award of a studentship. We thank Simon Gibbons and Vasily Belokurov
for many insightful discussions on the Sagittarius.

\bibliography{miynag}
\bibliographystyle{mn2e}

\end{document}